\documentclass[twocolumn,showpacs,preprintnumbers,amsmath,amssymb]{revtex4-2}
\usepackage{graphicx}
\usepackage{dcolumn}
\usepackage{bm}
\usepackage{lineno}
\newcommand{\bef}{\begin{figure}}
\newcommand{\eef}{\end{figure}}

\newcommand{\be}{\begin{equation}}
\newcommand{\ee}{\end{equation}}
\newcommand{\bea}{\begin{eqnarray}}
\newcommand{\eea}{\end{eqnarray}}
\begin{document}

\title{Anisotropic hadronic rescattering and its impact on $K^{*0}$ yield, and polarization observable}

\author{Kadambini Menduli,  Md. Nasim}
\affiliation{Department of Physical Sciences, Indian Institute of Science Education and Research, Berhampur, India}

\begin{abstract}
In this work, we investigate the anisotropic suppression of reconstructed $K^{*0}$ resonances arising from hadronic rescattering using the A Multi-Phase Transport (AMPT) model for Au+Au collisions at $\sqrt{s_{NN}}=200$ GeV. We demonstrate that the rescattering probability of the decay daughters depends strongly on the decay angle $\theta^{*}$ due to Lorentz boost effects, which lead to smaller laboratory-frame momenta for daughters emitted opposite to the parent particle motion. This anisotropic suppression influences several experimentally measured observables. We show that the reconstructed $K^{*0}$ yield exhibits a strong $\theta^{*}$ dependence. Furthermore, the anisotropic loss of resonances modifies the angular distributions used to extract the spin alignment parameter $\rho_{00}$ in the production-plane and helicity frames. Even in the absence of intrinsic polarization in the model, the reconstructed $K^{*0}$ sample shows deviations of $\rho_{00}$ from the unpolarized value of $1/3$, with opposite trends in the two reference frames. These results demonstrate that hadronic rescattering can generate apparent polarization signals and must be carefully considered in experimental measurements of vector-meson spin alignment using production plane and helicity frame. 

\end{abstract}
\maketitle

\section{Introduction}
\label{S:1}
Ultra-relativistic heavy-ion collisions provide a unique environment to study the properties of strongly interacting matter under extreme conditions of temperature, and energy density. Short-lived hadronic resonances provide a sensitive probe of the late-stage evolution of the fireball created in relativistic heavy-ion collisions~\cite{Brown_resonance, Markert_resonance, Schaffner, Rapp,star_resonance}. Among them, the $K^{*0}$ meson is particularly well suited for investigating hadronic phase dynamics because of its relatively short lifetime ($\sim 4$ fm/$c$), which is comparable to or shorter than the lifetime of the hadronic medium~\cite{system_life}. As a consequence, a significant fraction of $K^{*0}$ resonances decay inside the medium, making their measurable yields and kinematic properties sensitive to interactions occurring after chemical freeze-out.
The $K^{*0}$ predominantly decays via the hadronic channel $K^{*0}(\overline{K^{*0}})\rightarrow K^{\pm}\pi^{\mp}$ with a branching ratio of approximately $2/3$~\cite{pdg}. If the decay occurs within the dense hadronic environment, the daughter kaon and pion may undergo elastic or inelastic rescattering before reaching the detector. Such rescattering can alter their momenta or destroy the correlation between the decay products, thereby reducing the experimentally reconstructed $K^{*0}$ signal. Conversely, the large abundance of kaons and pions in the hadronic phase can lead to the reformation of the resonance through pseudo-elastic $K\pi \rightarrow K^{*0}$ processes, commonly referred to as regeneration~\cite{reco_issue_1,reco_issue_2,reco_issue_3,reco_issue_4}. The finally observed yield therefore reflects the competition between rescattering losses and regeneration gains during the hadronic stage.
A useful observable to quantify these competing effects is the $K^{*0}/K$ yield ratio. Since stable kaons are largely unaffected by late-stage hadronic interactions, deviations of this ratio from expectations based on chemical freeze-out conditions can signal medium-induced modifications. In particular, a decreasing $K^{*0}/K$ ratio with increasing system size or charged-particle multiplicity would indicate that rescattering dominates over regeneration, while a relative enhancement would point toward significant regeneration contributions.
Experimental measurements from various facilities~\cite{star_kstar_2002,star_kstar_2005,star_kstar_2008,star_kstar_2011,phenix_kstar_2014, NA49_kstar_2011, NA61_kstar_2020, NA61_kstar_2021,alice_kstar_2012,alice_kstar_2015,alice_kstar_2017,alice_kstar_2020_1,alice_kstar_2020_2,alice_kstar_2020_3,alice_kstar_2022,kstar_BES} consistently show that the $K^{*0}/K$ ratio in heavy-ion collisions is suppressed relative to that in smaller systems such as $p+p$ interactions. This suppression is commonly attributed to the dominance of hadronic rescattering effects in the extended and dense medium formed in central nucleus–nucleus collisions, highlighting the sensitivity of $K^{*0}$ production to the properties and lifetime of the hadronic phase.
Recent model calculations indicate that the anisotropic loss of $K^{*0}$ yield due to hadronic rescattering, relative to the reaction plane, can significantly modify the collective flow—particularly the directed flow—of the $K^{*0}$ resonance~\cite{parida}. In this paper, we investigate the anisotropic suppression of the $K^{*0}$ yield with respect to the momentum direction of the $K^{*0}$ itself and examine its consequences for the measured yield, and polarization observables.\\
 
\begin{figure}[ht]
\includegraphics[scale=0.4]{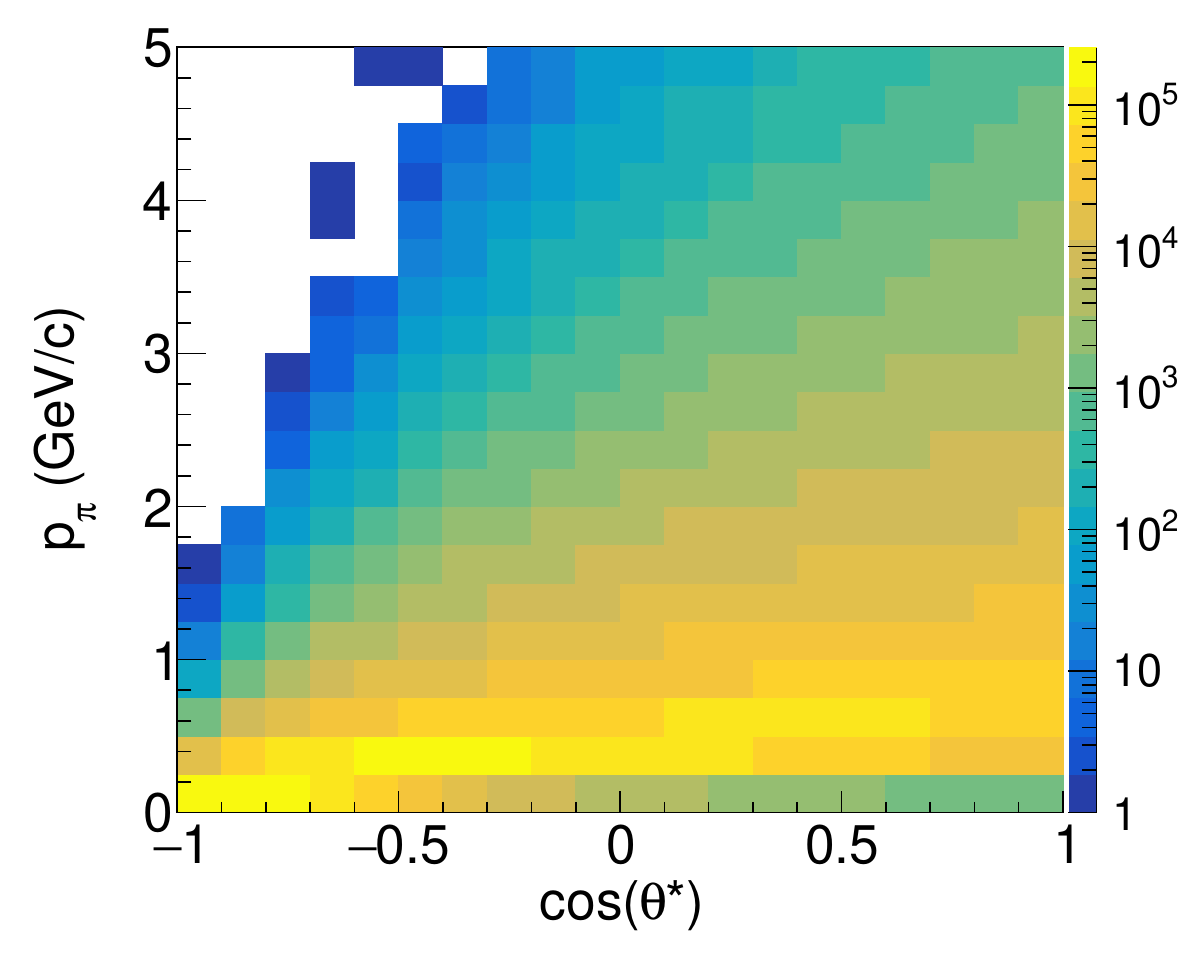}
\caption{The momentum distribution of pions from $K^{*0}$ decay as a function of cosine of decay angle $\theta^{*}$ obtained from AMPT model in Au+Au collisions at 200 GeV.}
\label{fig0}
\end{figure}

Let us consider a two-body decay of a parent particle (e.g. $K^{*0}$) moving along the $+z$ direction in the laboratory frame with velocity $\beta$ and Lorentz factor $\gamma = (1-\beta^2)^{-1/2}$. 
In the rest frame of the parents (denoted by $^*$), the four-momenta of the daughter are
\begin{equation}
p^{*\mu} = (E^*,\,  p^*),
\end{equation}
where $p^*$ is the magnitude of the three-momentum. If $\theta^{*}$ is the angle between the momentum of one of the daughters and the momentum of the parent particle in the Lab frame, then $p_z^* = p^*cos\theta^{*}$.

Under a Lorentz boost along the $z$ direction, the longitudinal momentum  the daughter emitted along ($\theta^{*} =0$) the parent motion ($p_z^* = +p^*$),
\begin{equation}
p_z^{\mathrm{(fwd)}} = \gamma (p^* + \beta E^*),
\end{equation}
while for the daughter emitted opposite ($\theta^{*} =\pi$) to the parent motion ($p_z^* = -p^*$),
\begin{equation}
p_z^{\mathrm{(bwd)}} = \gamma (-p^* + \beta E^*).
\end{equation}
The difference between the two longitudinal momenta is therefore
\begin{equation}
p_z^{\mathrm{(fwd)}} - p_z^{\mathrm{(bwd)}} 
= 2\gamma p^*.
\end{equation}
Since $\gamma > 0$ and $p^* > 0$, it follows that
\begin{equation}
p_z^{\mathrm{(fwd)}} > p_z^{\mathrm{(bwd)}}.
\end{equation}
Since a longitudinal Lorentz boost does not alter the transverse component of momentum, the daughter particle emitted opposite to the parent’s direction of motion acquires a smaller total momentum in the laboratory frame. Given that the probabilities of rescattering depend on the momentum of the particles, the rescattering effect becomes anisotropic with respect to the  angle $\theta^{*}$.

This paper is organised as follows. Section II briefly describes the AMPT model employed in this study. Section III presents the AMPT model results for the yields and  polarization observables  of $K^{*0}$ resonances . Finally, Section IV summarises the main findings and discusses the implications of this work for current experimental measurements in high-energy heavy-ion collisions at RHIC.

\section{AMPT Model}
The A Multi-Phase Transport (AMPT) model is a hybrid Monte Carlo framework developed to describe the space–time evolution of relativistic heavy-ion collisions from the initial nuclear impact to the final hadronic state~\cite{ampt}. The initial conditions are generated using HIJING~\cite{hijing}, which provides minijet partons from hard scatterings and excited strings from soft interactions. The subsequent partonic evolution is modeled through Zhang’s Parton Cascade (ZPC)~\cite{zpc}, where partons undergo two-body elastic scatterings with a cross section determined by perturbative QCD with a screening mass. AMPT is available in two configurations: the default version, in which partons recombine with strings and hadronize via Lund string fragmentation, and the string-melting version, where all excited strings are converted into partons and hadronization proceeds through quark coalescence. After hadronization, the hadronic interactions are simulated using the ART model~\cite{art}, which includes elastic and inelastic scatterings as well as resonance production and decay.
In this study, approximately one million events were generated for Au + Au 0-80\% minimum bias Au+Au collisions at RHIC energies. We have used string melting version of the AMPT in this study. Recent study shows that the AMPT model (string melting version) reasonably describes the measured $K^{*0}/K$ ratio in heavy-ion collisions at RHIC energies~\cite{barik}.  

\begin{figure*}[ht]
\begin{center}
\includegraphics[scale=0.4]{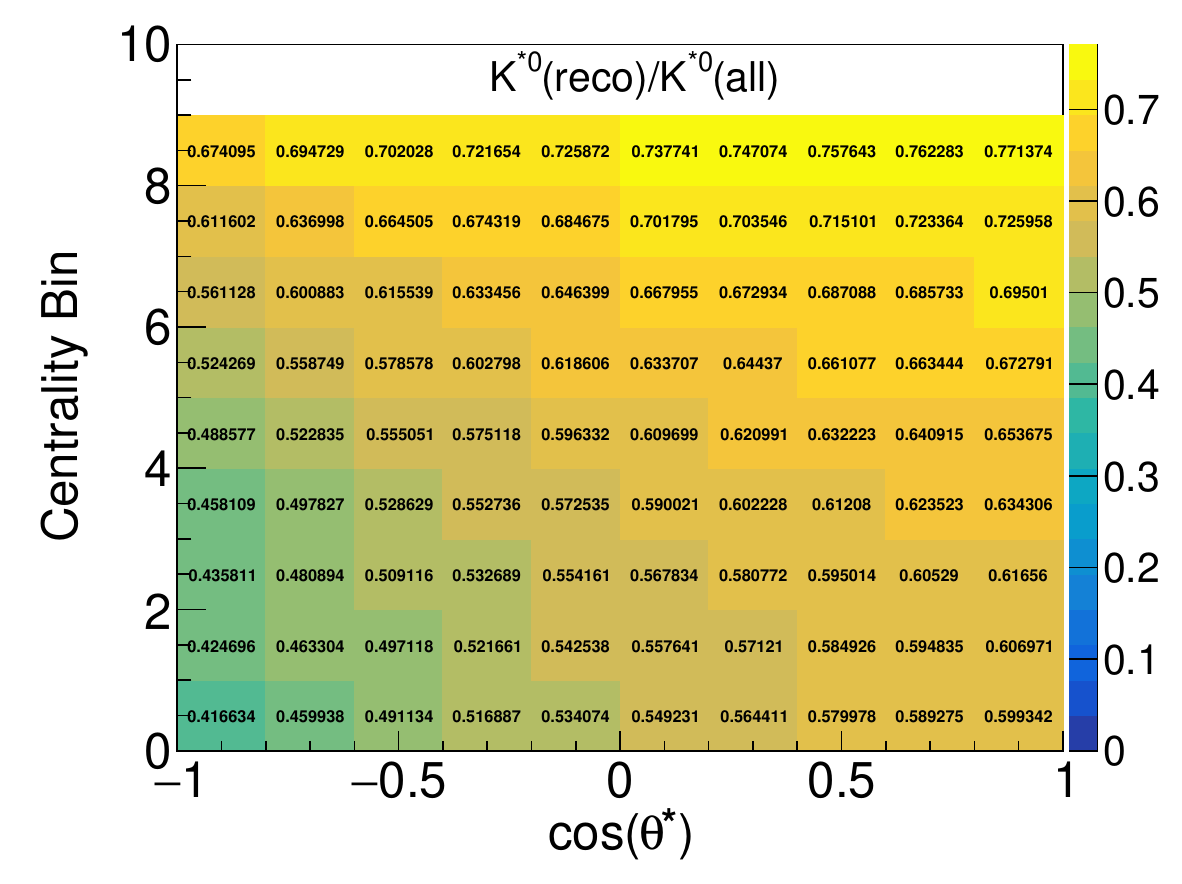}
\includegraphics[scale=0.4]{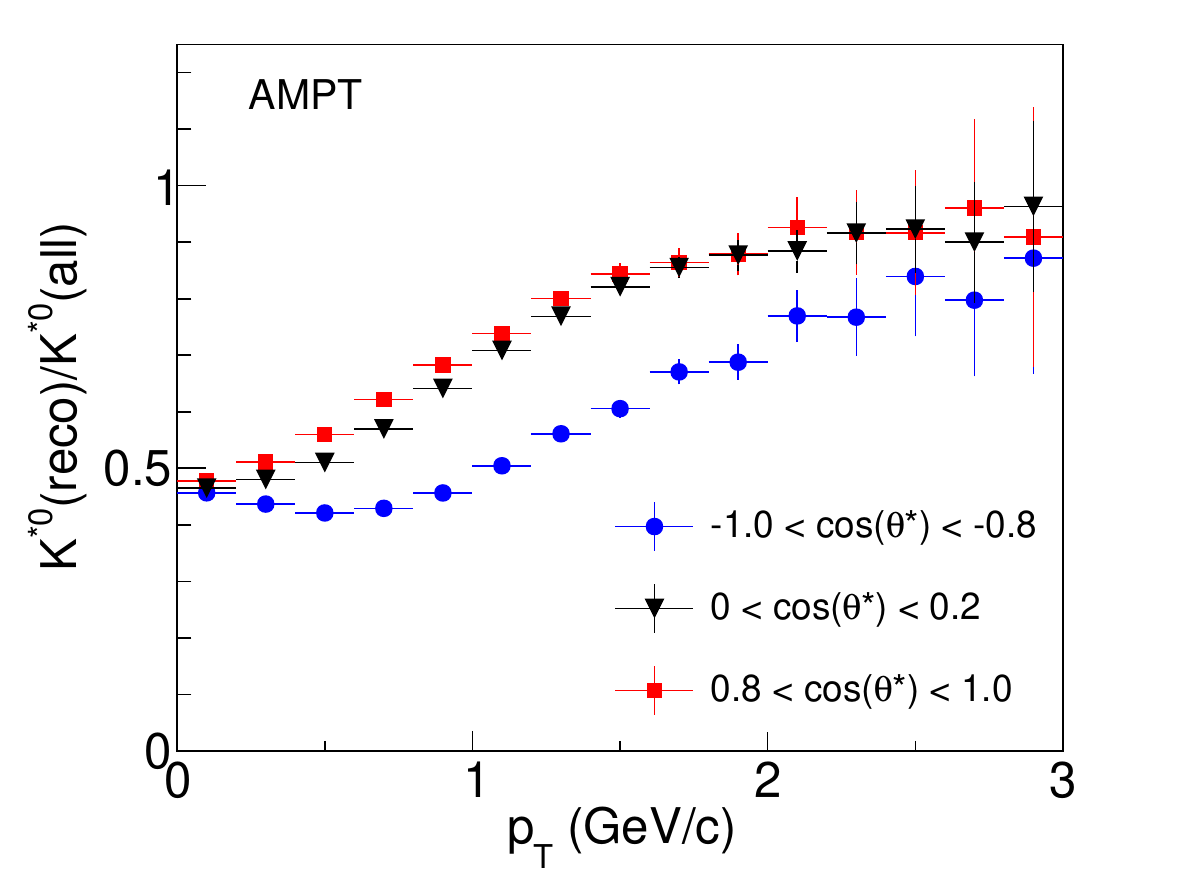}
\caption{Left panel: The ratio between reconstructed and all produced $K^{*0}$ for different centrality and $cos(\theta^{*})$. Right panel: The ratio between reconstructed and all produced $K^{*0}$ as function of transverse momentum  in 20-40\% centrality and for selected values of $cos(\theta^{*})$. This results is obtained using AMPT model (string melting version) in Au+Au collisions at $\sqrt{s_{NN}}$ =  200~GeV. }
\label{fig1}
\end{center}
\end{figure*}

\section{Result and DISCUSSION}
\subsection{Anisotropic Suppression of Yield}
The results are presented for both the total number of produced $K^{*0}$ mesons (denoted as $\mathrm{All}$) and those that can be reconstructed (denoted as $\mathrm{Reco}$) within the AMPT framework.
K$^{*0}$ resonances are reconstructed using the mass and momentum information of their decay daughters at the decay point. A K$^{*0}$ resonance is considered reconstructable only if both daughter particles leave the reaction zone without undergoing any further interactions, such that their momenta remain unchanged after the decay. Although this approach differs from the experimental method, in which K$^{*}$ resonances are identified through the invariant-mass spectra of $\pi$-K pairs, the theoretical reconstruction procedure based on tracking the decay daughters yields results consistent with those obtained using the experimental invariant-mass reconstruction technique. However, this theoretical method has an advantage over the invariant-mass reconstruction technique because it requires fewer events and allows $K^{*0}$ resonances to be identified directly in each event.
A detailed comparison of the two methods is presented in Appendix.



\begin{figure*}[ht]
\includegraphics[scale=0.5]{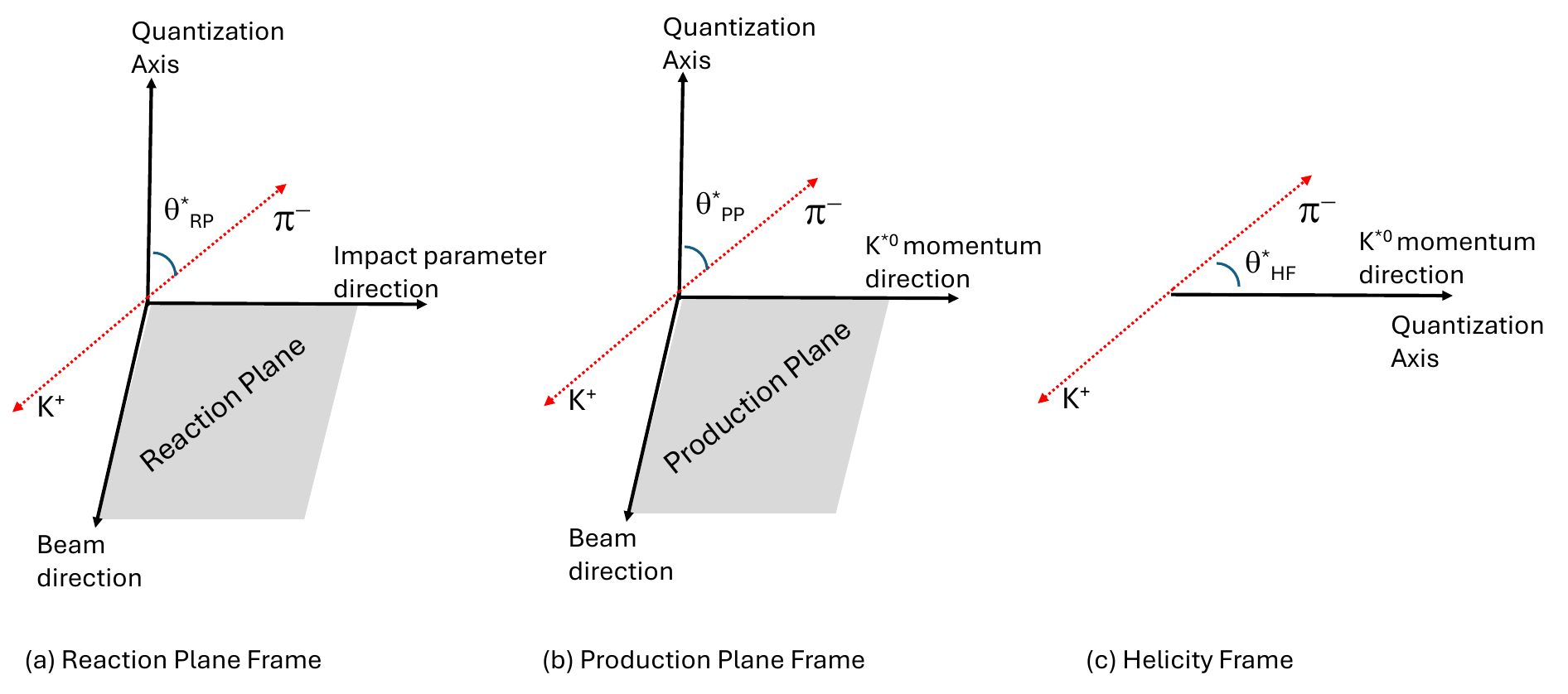}
\caption{ Three reference frames used to measured spin polarization: (1) the reaction plane frame, (2) the production plane frame, and (3) the helicity frame.  }
\label{fig_frame}
\end{figure*}

The momentum distribution of daughter pions from the  $K^{*0}$ decay as a function of the cosine of the decay angle $\theta^{*}$ is shown in Fig.~\ref{fig0}. This results is obatined using AMPT model in Au+Au collisions at 200GeV. We can see that the probability of low momentum particle is higher when pions emitted opposite to the parent’s direction of motion.

\begin{figure*}[ht]
\includegraphics[scale=0.8]{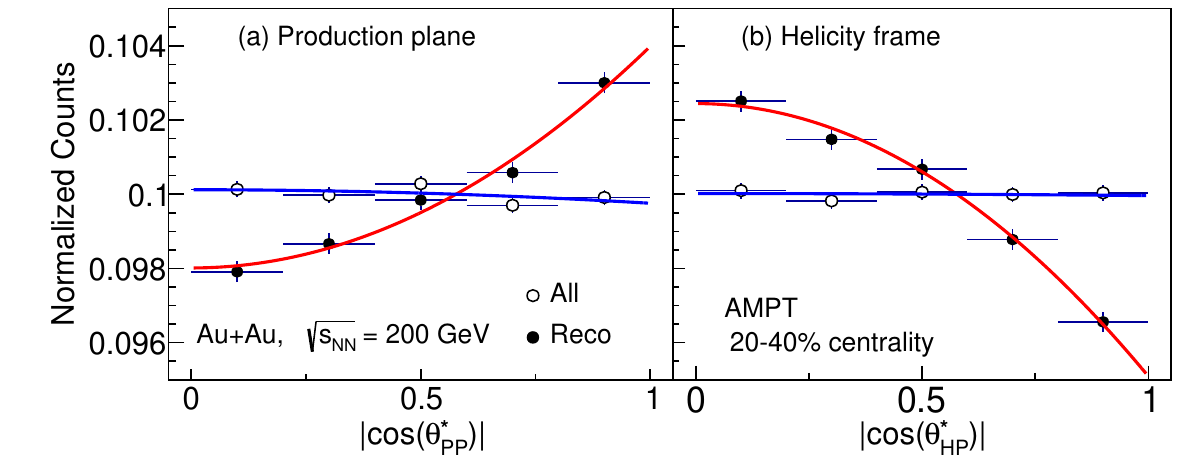}
\caption{ The yield distributions (normalized by total yield) of all produced and reconstructed $K^{*0}$ mesons as a function of $\cos(\theta^{*})$ with respect to the quantization axis defined in the production plane (PP) and helicity frame (HF). Red and blue curve are the fit to the data using Eq.~\ref{eq:rho00}. Results presented here are for 20-40\% Au+Au Collisions at at $\sqrt{s_{NN}}$ = 200 GeV using AMPT model.}
\label{fig_rho_fit}
\end{figure*}

The left panel of Fig.~\ref{fig1} shows the ratio of reconstructed to all produced $K^{*0}$ mesons as a function of the centrality bin and the decay angle $\theta^{*}$. Centrality bin 1 corresponds to the 0–5\% most central collisions, while bin 8 represents the 70–80\% most peripheral collisions. A pronounced variation of the ratio $K^{*0}(\mathrm{Reco})/K^{*0}(\mathrm{All})$ with the decay angle $\theta^{*}$ is observed for a given centrality bin. In particular, the ratio is smaller around $\theta^{*} \sim -1.0$ compared to that at $\theta^{*} \sim 1.0$. This behavior is consistent with the expectation that a daughter particle acquires a smaller total momentum in the laboratory frame; such low-momentum particles are therefore more susceptible to hadronic rescattering.
The transverse momentum dependence of the $K^{*0}(\mathrm{Reco})/K^{*0}(\mathrm{All})$ ratio is shown in the right panel of Fig.~\ref{fig1} for the 20–40\% centrality class and selected $\theta^{*}$ intervals. Figure~\ref{fig1} clearly demonstrates that the loss of yield due to rescattering is strongly anisotropic with respect to the decay angle $\theta^{*}$. Since the yield of reconstructed $K^{*0}$ mesons exhibits a strong dependence on $\theta^{*}$, the commonly measured yield ratio $K^{*0}/K$ is also expected to be sensitive to the decay angle. Therefore, yield measurements that probe rescattering effects (e.g., $K^{*0}/K$) should be performed as a function of $\theta^{*}$ in addition to their usual dependence on centrality and transverse momentum $p_{T}$. Such measurements would provide additional constraints on QCD-based transport models describing the hadronic phase of heavy-ion collisions.

%

\subsection{Polarization Observables}
Non-central heavy-ion collisions generate a large orbital angular momentum (OAM) in the system. A fraction of this OAM is transferred to the produced medium in the form of fluid vorticity aligned with the OAM direction, which can polarize particle spins via spin–orbit coupling, a phenomenon known as global polarization~\cite{spin_theory_prl}.
The spin alignment of a vector meson is described by a $3\times3$ Hermitian spin-density matrix with unit trace~\cite{spin_theory}. 
Neglecting the off-diagonal elements, the angular distribution of the decay products in the vector meson rest frame depends only on the diagonal matrix element $\rho_{00}$. 
The distribution can be expressed as
\begin{equation}
\frac{dN}{d\cos\theta^{*}} 
= N_{0}\left[(1-\rho_{00}) + (3\rho_{00}-1)\cos^{2}\theta^{*}\right],
\label{eq:rho00}
\end{equation}
where $N_{0}$ is a normalization constant and $\theta^{*}$ denotes the angle between the momentum of the decay daughter and the quantization axis.

\begin{figure}[ht]
\begin{center}
\includegraphics[scale=0.4]{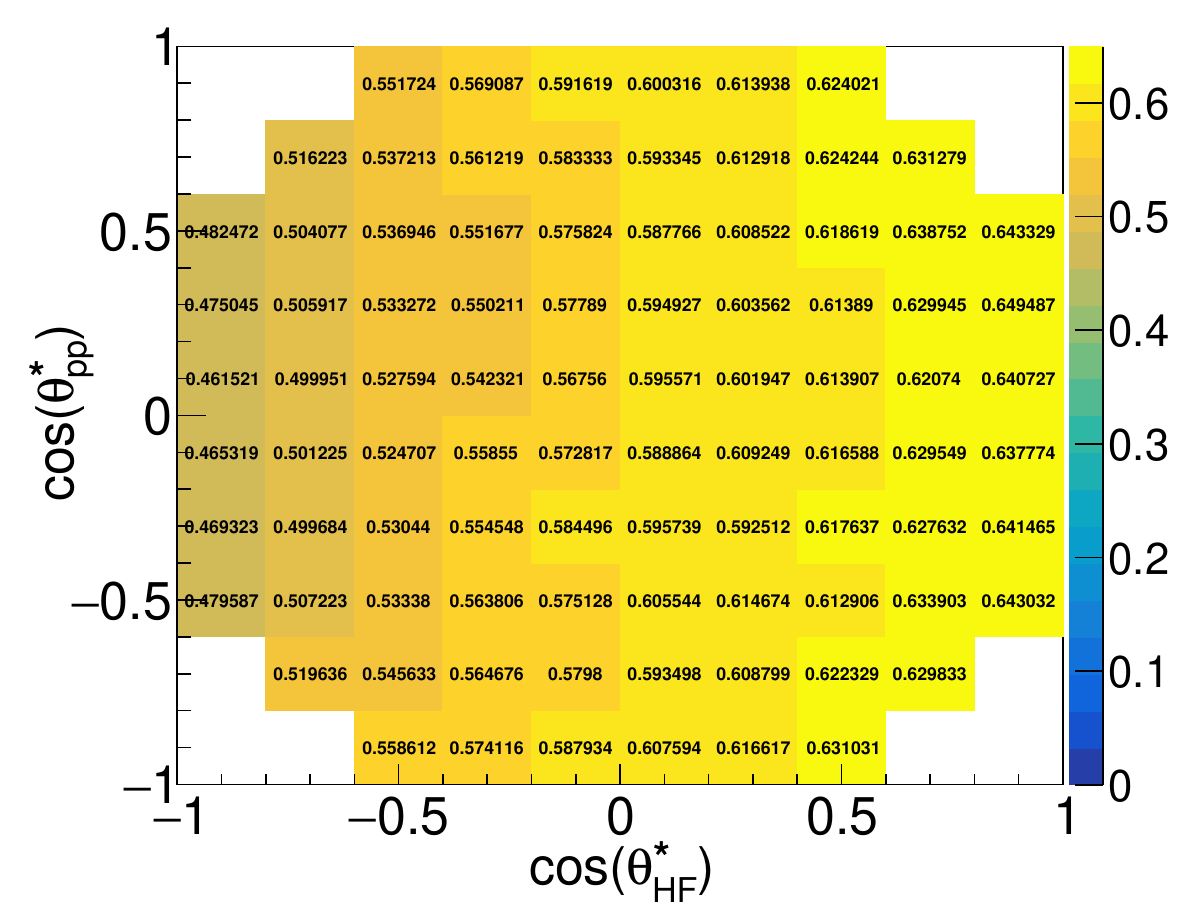}
\caption{$K^{*0}$ reconstruction probability as a function of $\cos(\theta^{*}_{PP})$ and $\cos(\theta^{*}_{HF})$ in Au+Au collision at $\sqrt{s_{NN}}$ = 200 GeV using AMPT model.}
\label{fig_reco_prob}
\end{center}
\end{figure}

In spin alignment studies of vector mesons, three reference frames (see Fig.~\ref{fig1}) are used to choose the quantization axis: (1) the reaction plane frame~\cite{spin_star_200,spin_star_bes}, (2) the production plane frame~\cite{spin_star_200,spin_star_bes}, and (3) the helicity frame~\cite{spin_delphi,spin_opal}. Each frame provides a different direction relative to which the angular distribution of the decay particles is measured.
In the reaction plane frame, the quantization axis is taken perpendicular to the reaction plane. The reaction plane is defined by the beam direction and the impact parameter. This axis represents the direction of the system’s global angular momentum in non-central collisions.
In the production plane frame, the quantization axis is defined as perpendicular to the plane formed by the beam direction and the meson’s momentum.
In the helicity frame, the quantization axis is chosen along the meson’s momentum in the laboratory frame. The angle $\theta^*$ is measured between this axis and the momentum of one of the decay particles in the meson’s rest frame.\\

\begin{figure*}[ht]
\includegraphics[scale=0.8]{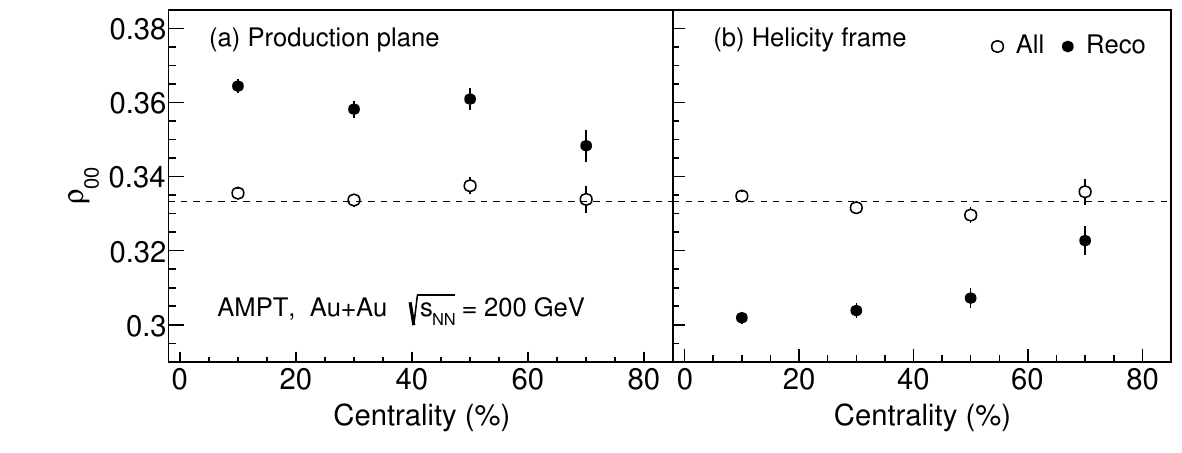}
\caption{ The extracted values of $\rho_{00}$ using Eq.~\ref{eq:rho00} as function of collisions centrality for both all produced and reconstructed $K^{*0}$ mesons in the (a) production-plane and (b) helicity frames.}
\label{fig_rho_cen}
\end{figure*}

The global polarization of vector mesons, such as the $\phi$ and $K^{*0}$, has been measured at both RHIC~\cite{spin_star_200,spin_star_bes} and LHC energies~\cite{spin_alice}. Owing to its short lifetime, the $K^{*0}$ meson is particularly sensitive to hadronic interactions, and the probability of rescattering exhibits anisotropy with respect to the chosen quantization axis. As demonstrated in this work, the rescattering probability depends strongly on the direction of the parent particle momentum. Consequently, the measurement of the spin alignment parameter $\rho_{00}$ for short-lived resonances such as the $K^{*0}$ can be significantly affected when evaluated in frames where the quantization axis is defined by the momentum direction of the $K^{*0}$, such as the production plane and helicity frames.
The influence of hadronic rescattering on the spin alignment of the $K^{*0}$ has previously been investigated using the AMPT model in the reaction plane frame~\cite{ampt_rho_00}. In the present study, we examine the sensitivity of hadronic rescattering effects when the spin alignment is measured in the production plane and helicity frames.

\begin{figure}
\includegraphics[scale=0.35]{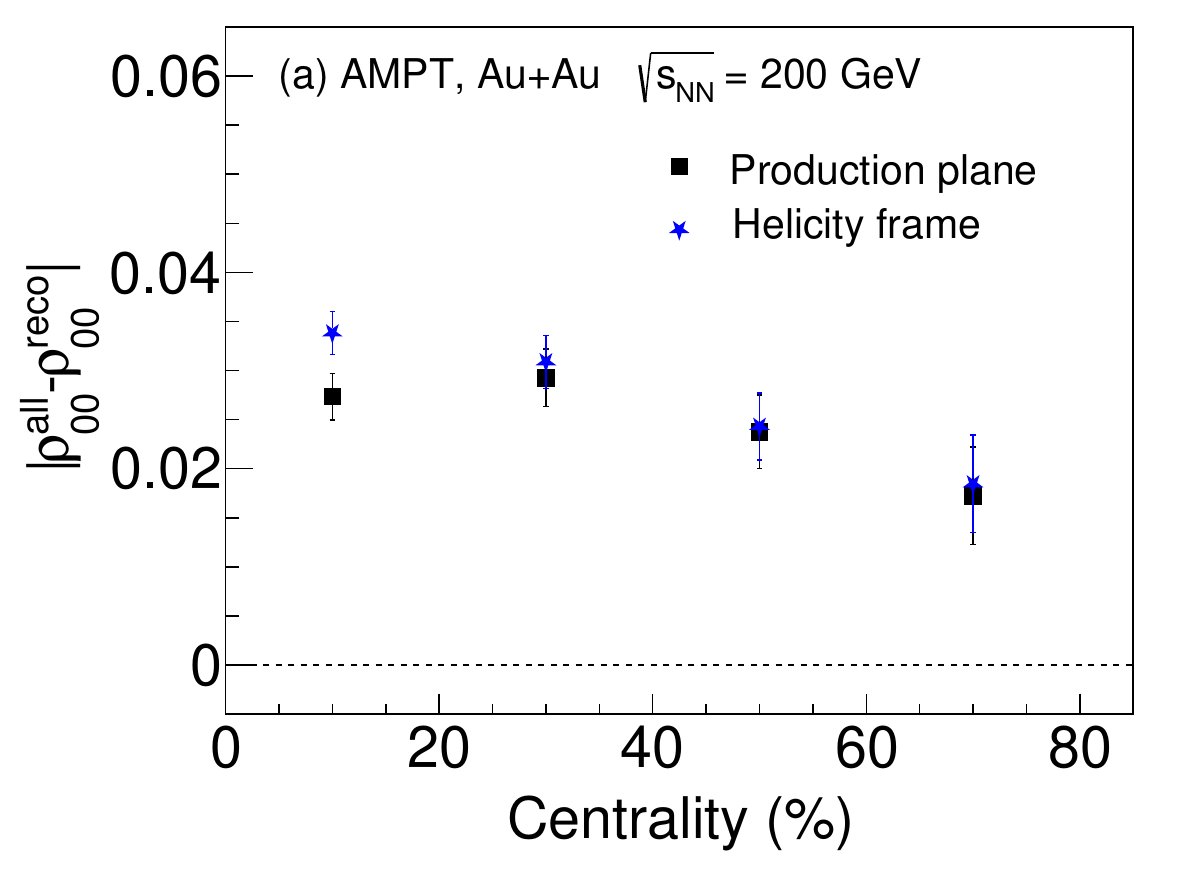}
\includegraphics[scale=0.35]{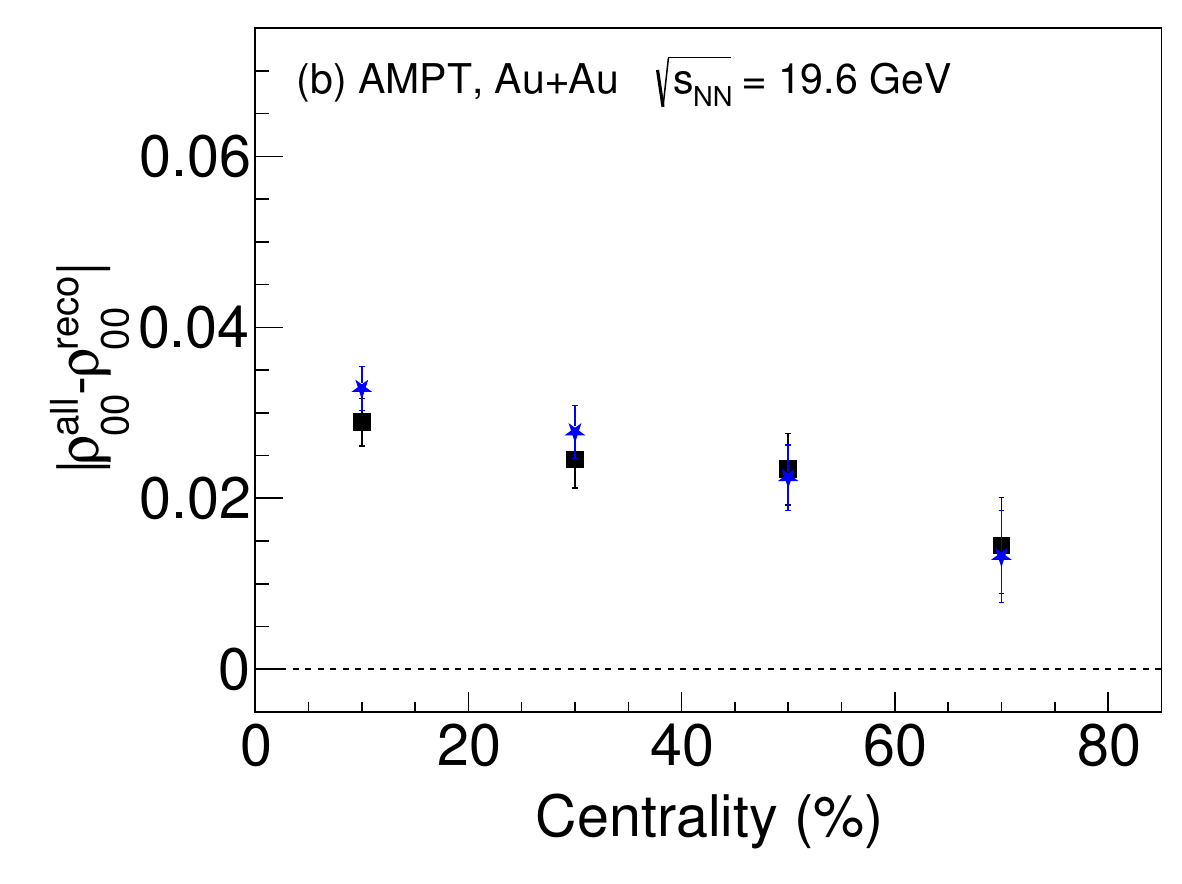}
\includegraphics[scale=0.35]{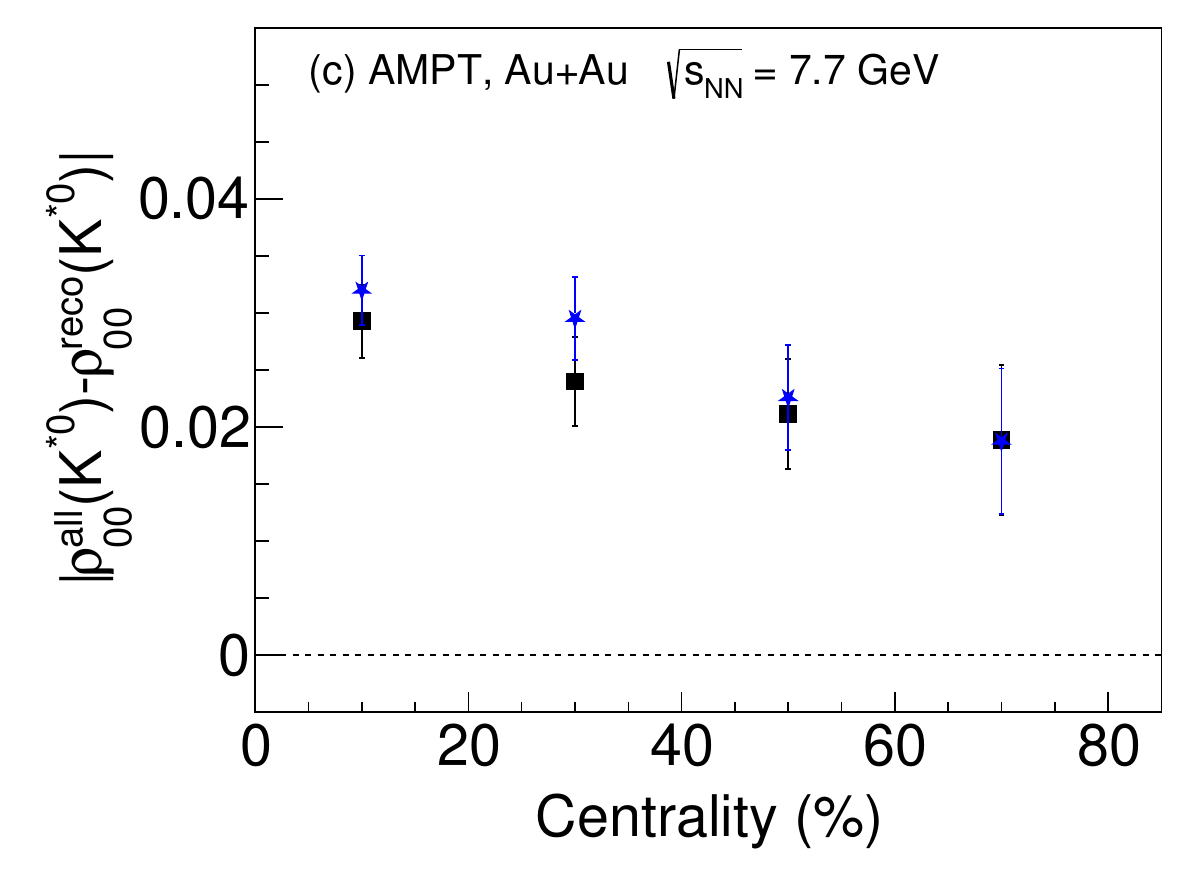}
\caption{The absolute values of the difference in magnitude of $\rho_{00}$ from 1/3 due to rescattering measured using the production plane and the helicity frame using the AMPT model.
Panel (a), (b) and (c) represent calculation done at 200 GeV, 19.6 and 7.7 GeV, respectively. }
\label{fig_delta_rho_cen}
\end{figure}

Figure~\ref{fig_rho_fit} shows the yield distributions of all produced and reconstructed $K^{*0}$ mesons as a function of $\cos(\theta^{*})$ with respect to the quantization axis defined in the production plane (PP) and helicity frame (HF). The calculation is performed using the string-melting version of the AMPT model for Au+Au collisions at $\sqrt{s_{NN}} = 200$~GeV. The yield distribution of all produced $K^{*0}$ mesons is nearly flat, since spin-polarization effects are not implemented in the AMPT model used in this study. Fitting this distribution using Eq.~\ref{eq:rho00} yields a value of $\rho_{00}$ very close to $1/3$, as shown in Fig.~\ref{fig_rho_cen}.
In contrast, the yield distribution of reconstructed $K^{*0}$ mesons shows a significant modification as a function of $\cos(\theta^{*}_{PP})$ and $\cos(\theta^{*}_{HF})$. This modification arises from an asymmetric loss of $K^{*0}$ yield with respect to the decay angles $\theta^{*}_{PP}$ and $\theta^{*}_{HF}$. The yield loss is largest when $\cos(\theta^{*}_{PP}) \sim 0$ and $\cos(\theta^{*}_{HF}) \sim 1$.
This behavior can be understood from the kinematics of the decay daughters. As discussed earlier, the daughter particle emitted opposite to the $K^{*0}$ momentum has a smaller momentum in the laboratory frame, making it more susceptible to rescattering. Consequently, the rescattering probability is increased when $\theta^{*}_{HF}$ is close to $180^\circ$ ($|\cos(\theta^{*}_{HF})| \sim 1$). In the production-plane frame, the polarization axis is perpendicular to the $K^{*0}$ momentum, leading to increased rescattering when $\theta^{*}_{PP}$ is close to $90^\circ$ ($|\cos(\theta^{*}_{PP})| \sim 0$).

\begin{figure}
\includegraphics[scale=0.35]{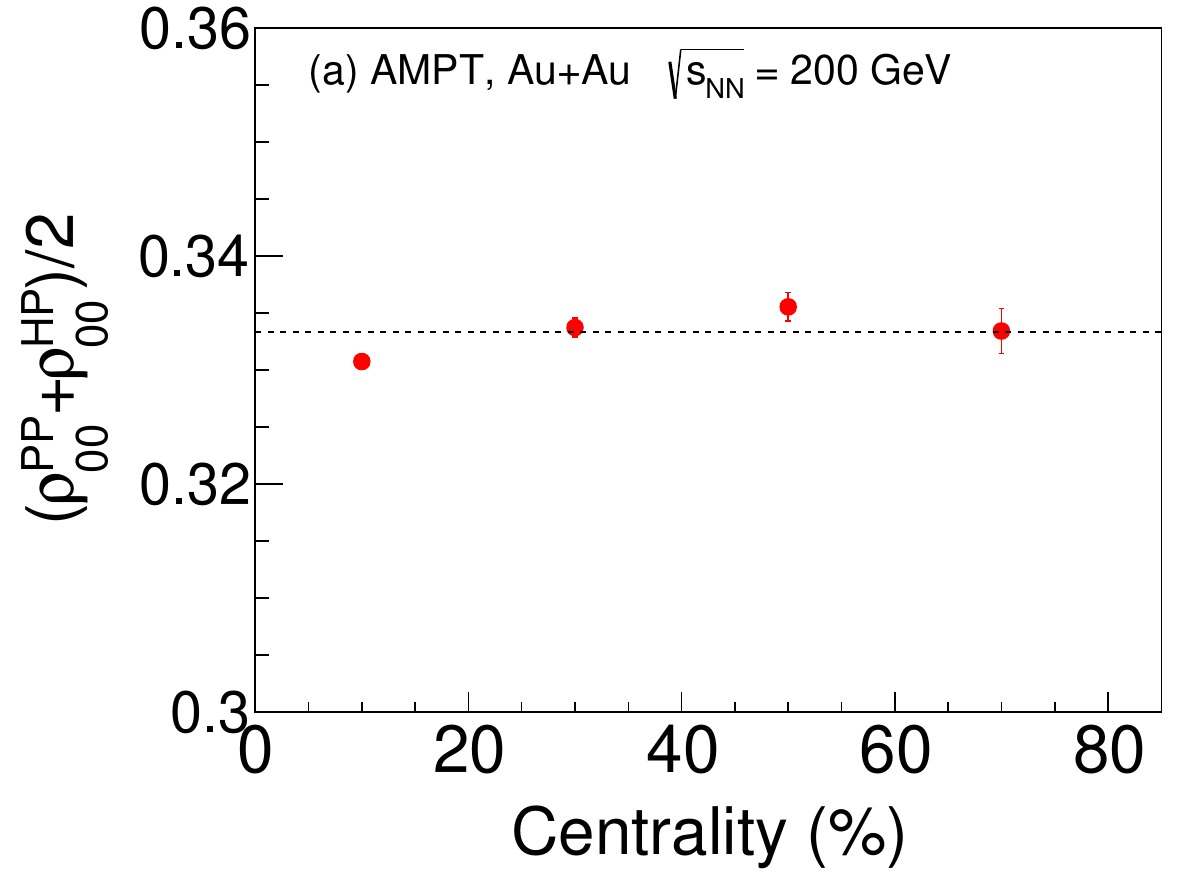}
\includegraphics[scale=0.35]{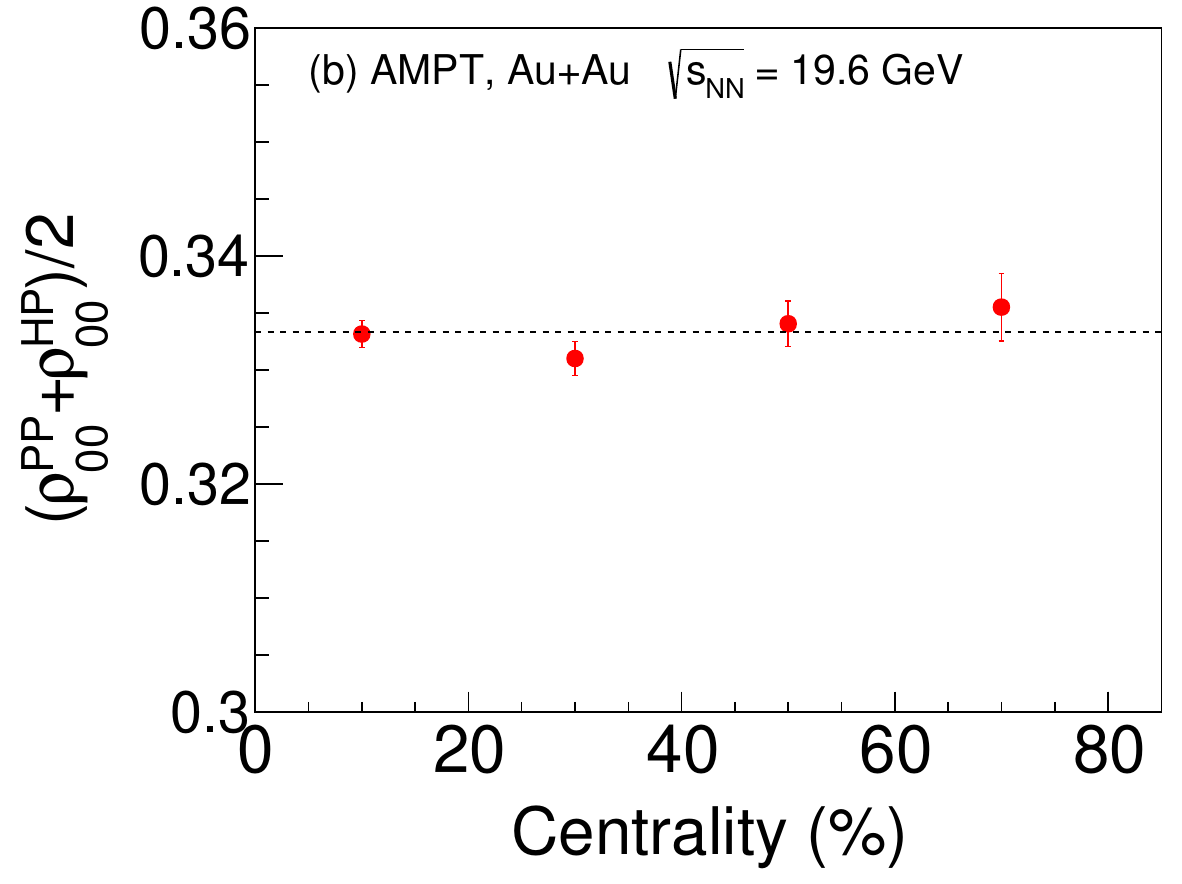}
\includegraphics[scale=0.35]{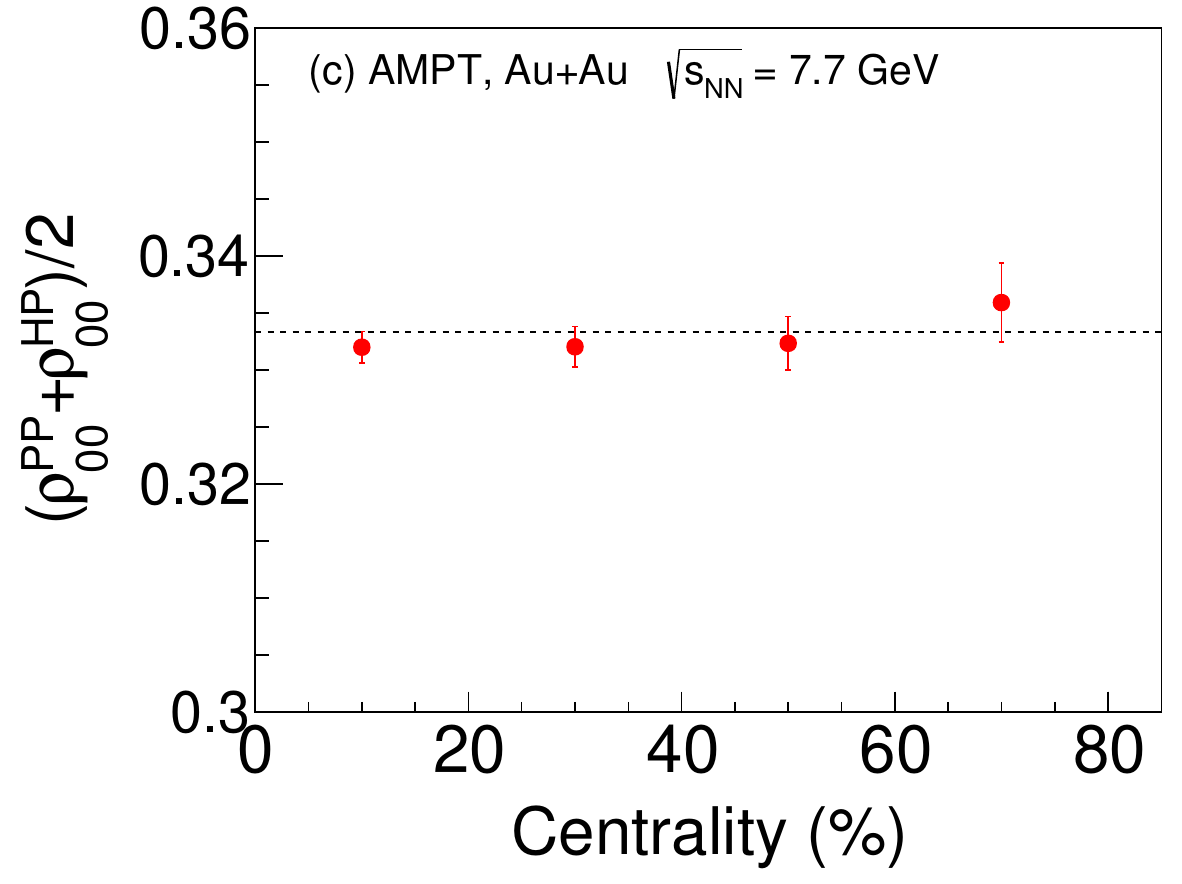}
\caption{The average values $\rho_{00}$ measured in the production plane and helicity frame using the AMPT model.
Panel (a), (b) and (c) represent calculation done at 200 GeV, 19.6 and 7.7 GeV, respectively. }
\label{fig_avg_rho_cen}
\end{figure}

Figure~\ref{fig_reco_prob} presents the $K^{*0}$ reconstruction probability as a function of $\cos(\theta^{*}_{PP})$ and $\cos(\theta^{*}_{HF})$. The reconstruction probability is lowest near $\cos(\theta^{*}_{PP}) \sim 0$ and $\cos(\theta^{*}_{HF}) \sim -1$. For a fixed value of $\cos(\theta^{*}_{PP})$, the reconstruction probability changes significantly with $\cos(\theta^{*}_{HF})$. In contrast, for a fixed $\cos(\theta^{*}_{HF})$, the variation with $\cos(\theta^{*}_{PP})$ is relatively small.
Figure~\ref{fig_rho_cen} shows the extracted values of $\rho_{00}$ for reconstructed $K^{*0}$ mesons in both the production-plane and helicity frames. Due to hadronic rescattering effects, the observed values of $\rho_{00}$ deviate from $1/3$, even though no polarization mechanism is implemented in the AMPT model. It is noteworthy that the extracted $\rho_{00}$ values are larger than $1/3$ in the production-plane frame, while they are smaller than $1/3$ in the helicity frame.

Finally, we compare the change in the magnitude of $\rho_{00}$ arising from the re-scattering effect using two methods: the production plane and the helicity frame. Figure~\ref{fig_delta_rho_cen} presents the absolute deviation of $\rho_{00}$ from $1/3$ due to rescattering, evaluated in the production plane and helicity frame within the AMPT model. The results are shown as a function of collision centrality.

Panel (a) of Fig.~\ref{fig_delta_rho_cen} displays the calculation using the string melting version of the AMPT model at $\sqrt{s_{NN}} = 200$ GeV. Interestingly, the observed deviation of $\rho_{00}$ from $1/3$ is nearly identical for both the production plane and helicity frame methods. To further verify this observation, we performed the same analysis at $\sqrt{s_{NN}} = 19.6$ GeV (shown in panel (b)) and at $\sqrt{s_{NN}} = 7.7$ GeV (shown in panel (c)). In all cases, the deviation of $\rho_{00}$ from $1/3$ remains nearly the same for the production plane and the helicity frame methods, but in the opposite direction. Therefore, taking the average of $\rho_{00}$ values measured in the production plane and the helicity frame will largely cancel out the effect of hadronic re-scattering and the magnitude of $\rho_{00}$ will be close to 1/3, as shown in Fig.~\ref{fig_avg_rho_cen}. It would be interesting to investigate this behavior using models that incorporate finite polarization effects, which are beyond the scope of the present study since no such models are available as open source. 

\section{Summary}
In this work, we investigate the effect of anisotropic hadronic rescattering on the measurable properties of the $K^{*0}$ resonance in relativistic heavy-ion collisions using the AMPT transport model. Due to Lorentz boost kinematics, the daughter particle emitted opposite to the parent momentum acquires a smaller momentum in the laboratory frame and is therefore more susceptible to hadronic rescattering. As a result, the probability of reconstructing $K^{*0}$ mesons from their decay products becomes strongly dependent on the decay angle $\theta^{*}$.
AMPT calculations for Au+Au collisions at RHIC energies show that the ratio of reconstructed to all produced $K^{*0}$ mesons exhibits a pronounced dependence on $\cos(\theta^{*})$, indicating anisotropic yield loss caused by rescattering.
We further studied the impact of this effect on spin alignment measurements of the $K^{*0}$. The angular distributions of reconstructed resonances are significantly distorted relative to those of all produced resonances, leading to deviations of the extracted spin-alignment parameter $\rho_{00}$ from the unpolarized value of $1/3$, even in the absence of intrinsic polarization in the model. The reconstructed $K^{*0}$ sample shows $\rho_{00} > 1/3$ in the production-plane frame and $\rho_{00} < 1/3$ in the helicity frame, but the deviation from 1/3 is nearly equal in both the method. As a result, averaging $\rho_{00}$ the values obtained from these two frames largely cancels the effect of hadronic rescattering. It is therefore important to verify this observation using a model that incorporates finite polarization.

\section{Appendix}
This section provides additional details on the reconstruction of $K^{*0}$ resonances used in this work and compares it with the method commonly employed in experimental data analyses.
Experimentally, the $K^{*0}$ signal is reconstructed from the invariant-mass distribution of oppositely charged kaon--pion pairs formed within the same event (unlike-sign pairs). This distribution contains the $K^{*0}$ signal peak superimposed on a large combinatorial background arising from uncorrelated kaon--pion pairs. The combinatorial background is typically estimated using either the mixed-event technique or the track-rotation method.
In the mixed-event technique, kaons and pions from different events with similar global characteristics are combined to reproduce the distribution of uncorrelated pairs. In the track-rotation method, the momentum vector of one of the daughter particles (in this case, the pion) is rotated by $180^{\circ}$ in the transverse plane, thereby removing the correlation between daughter particles originating from the same parent resonance while preserving the single-particle kinematics. In the present study, the track-rotation method is used, following the procedure used in the recent publication of the STAR collaboration~\cite{kstar_BES,bes2_kstar}. We refer to this approach as the \emph{Experimental Method}.

\begin{figure*}[ht]
\includegraphics[scale=0.35]{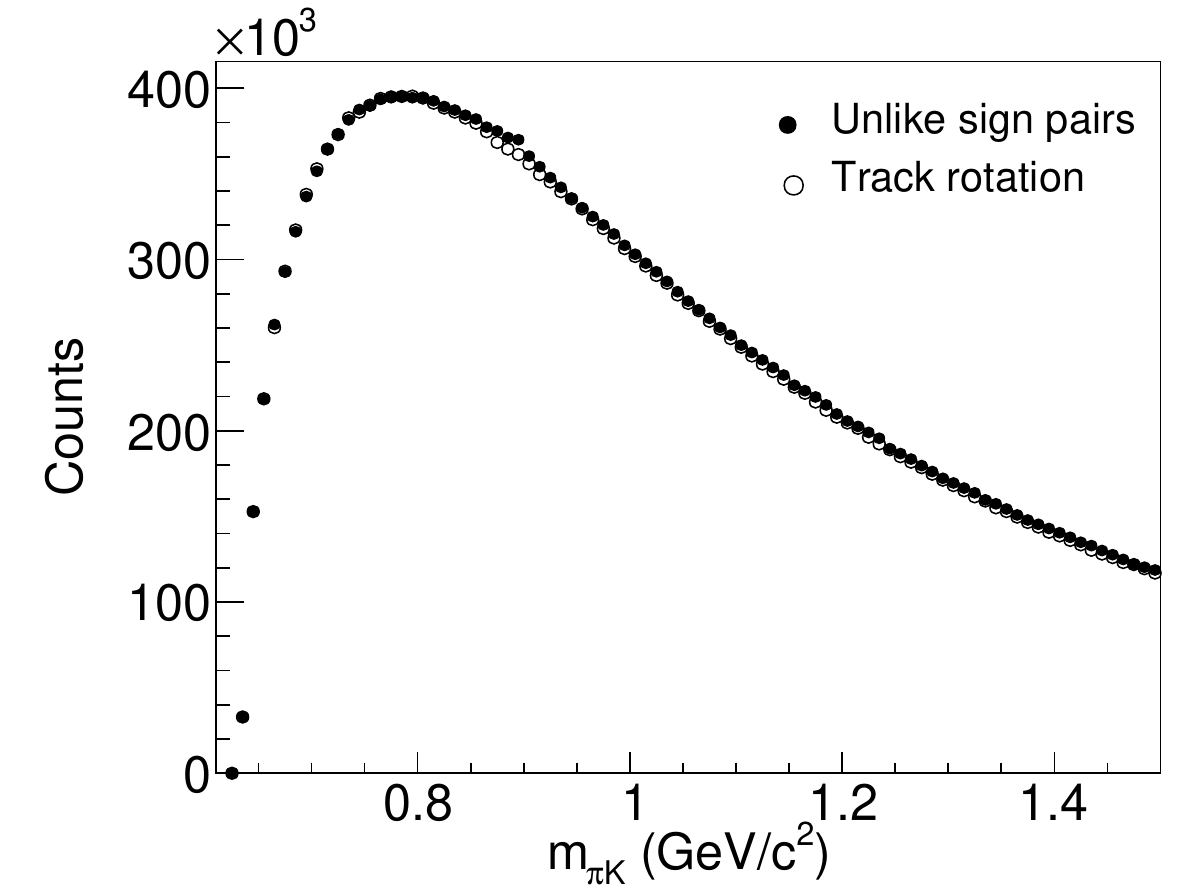}
\includegraphics[scale=0.35]{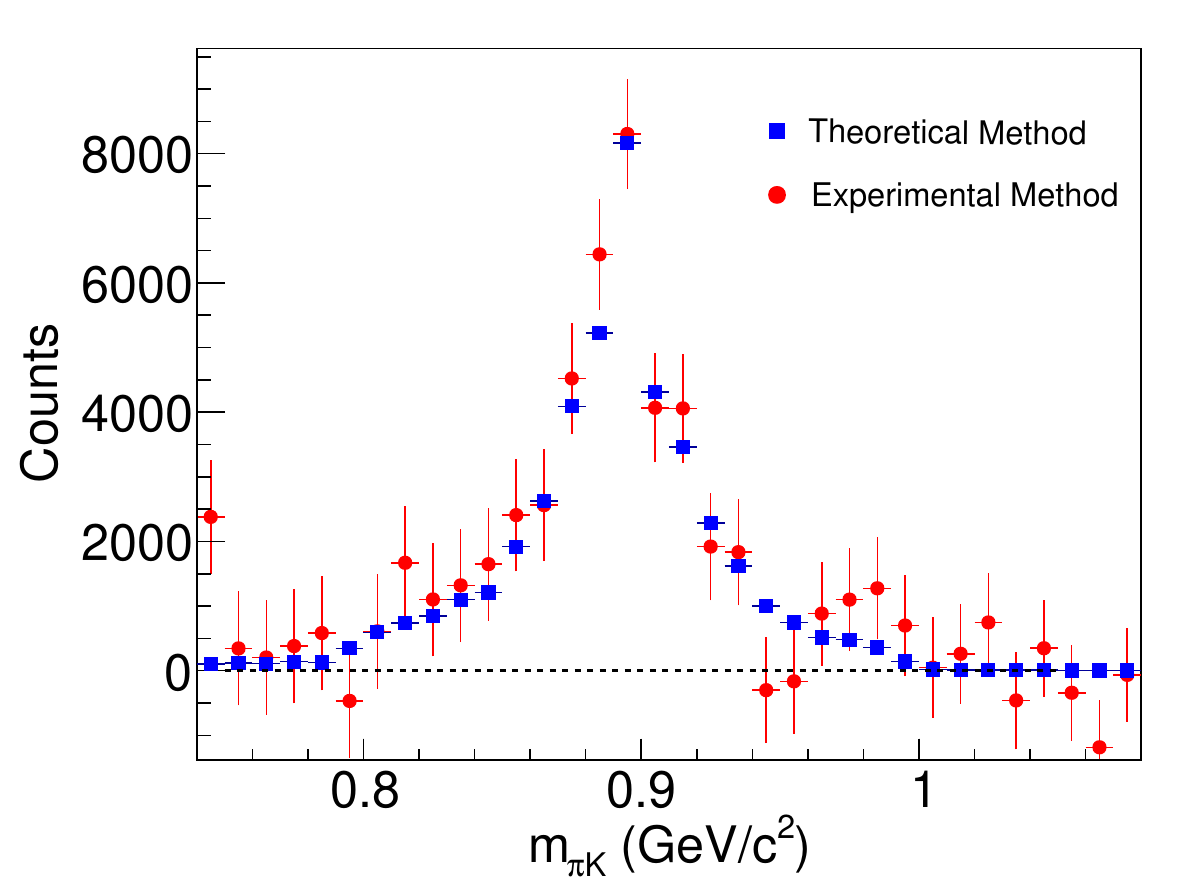}
\caption{Left panel: The invariant-mass distribution of unlike-sign kaon--pion pairs together with the combinatorial background estimated using the track-rotation method in Au+Au collisions at 19.6 GeV using AMPT model. Right panel: The combinatorial background-subtracted $K^{*0}$ signal. For comparison, the $K^{*0}$ yield obtained using the  \emph{Theoretical Method} is also shown.  }
\label{app_fig}
\end{figure*}

In contrast, the reconstruction procedure in the present model study is based on directly tracking the decay daughters of the $K^{*0}$ resonance. A $K^{*0}$ resonance is considered reconstructable only if both daughter particles escape the reaction zone without undergoing any further interactions. In such cases, the original resonance is counted as observable. This procedure is referred to as the \emph{Theoretical Method}.\\
Figure~\ref{app_fig} (left panel) shows the invariant-mass distribution of unlike-sign kaon--pion pairs together with the combinatorial background estimated using the track-rotation method. The corresponding background-subtracted $K^{*0}$ signal is shown in the right panel of Fig.~\ref{app_fig}. For comparison, the $K^{*0}$ yield obtained using the  \emph{Theoretical Method} is also shown. The results obtained from the two methods are found to be consistent within the estimated statistical uncertainties, demonstrating the validity of the theoretical reconstruction procedure adopted in this work. Our finding is consistent with the previous study using UrQMD model~\cite{urqmd_kstar}.

\begin{figure*}[ht]
\includegraphics[scale=0.35]{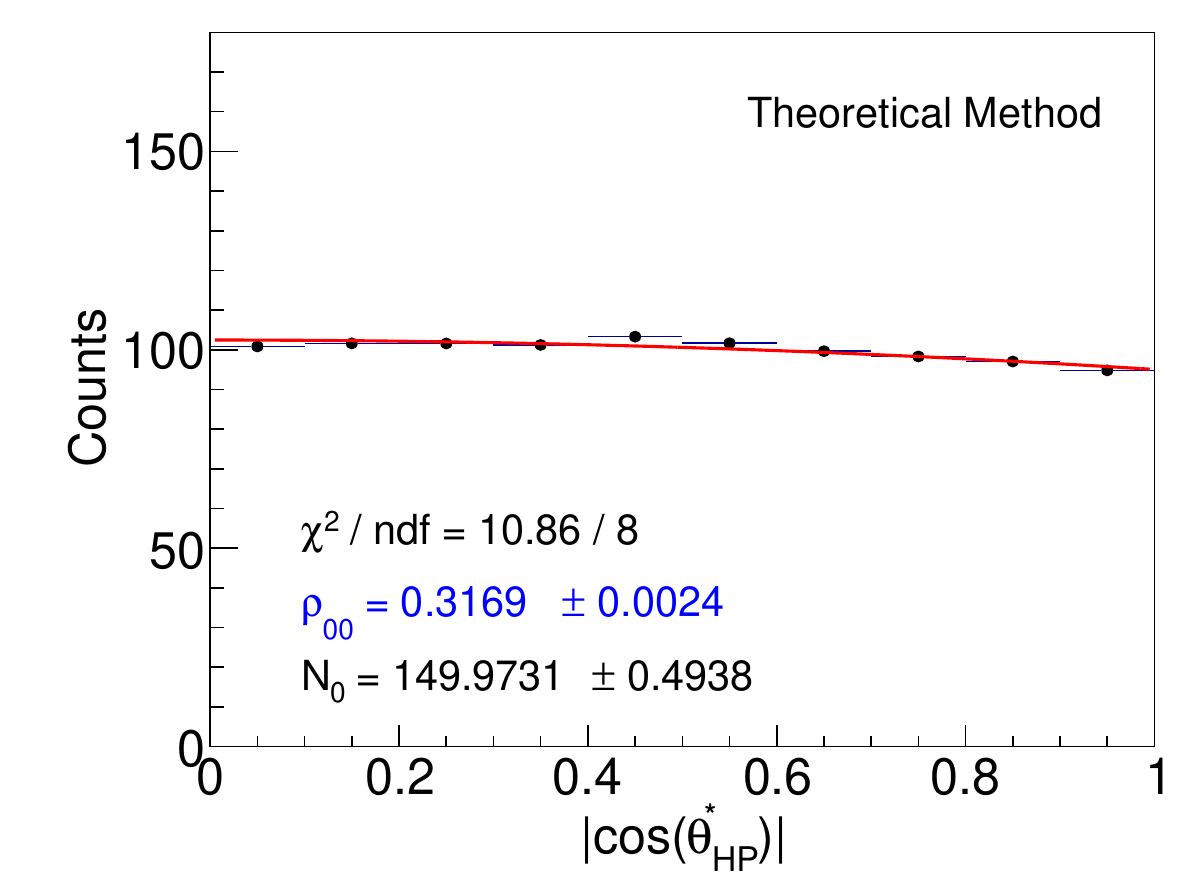}
\includegraphics[scale=0.35]{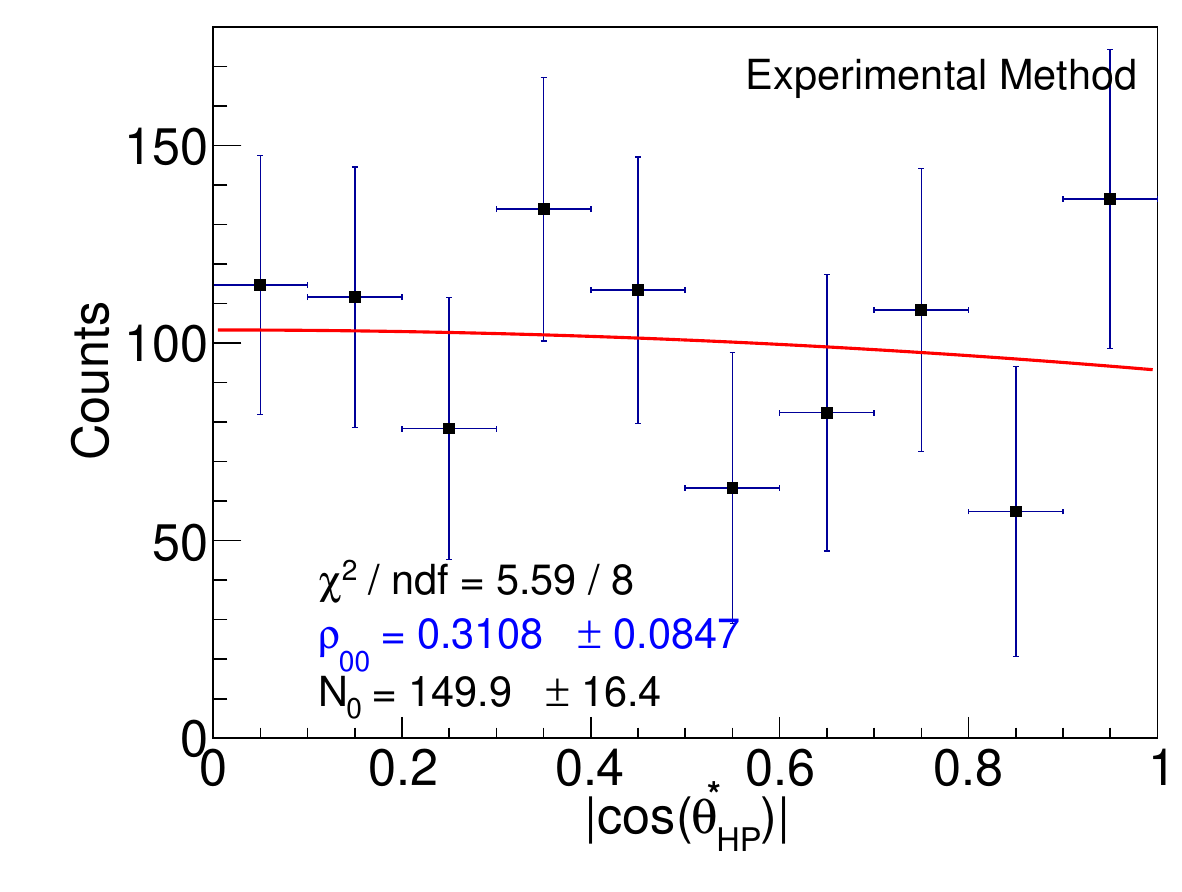}
\caption{The yield distributions  of  reconstructed $K^{*0}$ mesons using both \emph{Theoretical Method} (left panel) and \emph{Experimental Method} (right panel) as a function of $\cos(\theta^{*})$ with respect to the quantization axis defined in the helicity frame (HF). Red  curves are the fit to the data using Eq.~\ref{eq:rho00}. Results presented here are for 40-60\% Au+Au collisions at at $\sqrt{s_{NN}}$ = 19.6 GeV using AMPT model.  }
\label{app_fig_rho}
\end{figure*}

Finally, we have shown a comparison of spin-alignment parameter $\rho_{00}$ extracted by fitting  the yield of reconstructed $K^{*0}$ mesons using both \emph{Theoretical Method} (left panel of Fig.\ref{app_fig_rho}) and \emph{Experimental Method} (right panel of Fig.\ref{app_fig_rho}) as a function of $\cos(\theta^{*})$ with respect to the quantization axis defined in the helicity frame (HF). We can see from Fig.\ref{app_fig_rho} that both method gives consistent values of $\rho_{00}$ within statistical uncertainties which further demonstrate the validity of the theoretical reconstruction procedure used in this simulation work.

%

\end{document}